\pdfoutput=1
\documentclass[pre,superscriptaddress,showpacs,twocolumn]{revtex4}

\usepackage{graphicx}
\usepackage{appendix}
\usepackage{graphicx,amsfonts}
\usepackage{epsfig,amsmath}
\usepackage{verbatim}
\usepackage{color}

\def\be{\begin{equation}}
\def\ee{\end{equation}}
\def\bea{\begin{eqnarray}}
\def\eea{\end{eqnarray}}

\begin{document}

\title{Short time growth of a KPZ interface with flat initial conditions}

\author{Thomas Gueudr\'e}
\affiliation{CNRS-Laboratoire de Physique
Th{\'e}orique de l'Ecole Normale Sup{\'e}rieure, 24 rue Lhomond, 75231
Paris Cedex, France}
\author{Pierre Le Doussal}
\affiliation{CNRS-Laboratoire de Physique
Th{\'e}orique de l'Ecole Normale Sup{\'e}rieure, 24 rue Lhomond, 75231
Paris Cedex, France}
\author{Alberto Rosso}
\affiliation{CNRS-Universit\'e Paris-Sud, LPTMS, UMR8626-B\^at 100,91405 Orsay Cedex, France}
\author{Adrien Henry}
\affiliation{CNRS-Universit\'e Paris-Sud, LPTMS, UMR8626-B\^at 100,91405 Orsay Cedex, France}
\author{Pasquale Calabrese}
\affiliation{Dipartimento di Fisica dell'Universit\`a di Pisa and INFN, 56127 Pisa Italy}

\date{\today}

\begin{abstract}
The short time behavior of the 1+1 dimensional KPZ growth equation with a flat initial condition is obtained from
the exact expressions 
for the moments of the partition function of a directed polymer  with one endpoint  free and the other fixed. 
From these expressions, the short time expansions of the lowest cumulants of the KPZ height field are exactly derived. 
The results for these two classes of cumulants are checked in high precision lattice numerical simulations. 
The short time limit considered here is relevant for the study of the interface growth in the large diffusivity/weak noise limit, 
and describes the {\it universal} crossover between the Edwards-Wilkinson and KPZ universality classes
for an initially flat interface. 
\end{abstract}


\maketitle


\section{Introduction}

The growth of interfaces in the presence of noise can be classified into several universality classes. 
When the growth rate does not depend on the slope of the interface, the growth process falls in the simplest 
Edwards-Wilkinson (EW) class. 
When instead the growth rate is slope dependent (e.g. in the so-called lateral growth), the process falls into the
class defined by the non-linear continuum Kardar-Parisi-Zhang (KPZ) equation \cite{KPZ,kpzreviews}
\begin{align}
\partial _t h = \nu \nabla ^2 h +\frac{1}{2} \lambda _0 (\nabla h)^2 + \xi (x,t),
\label{kpzeq}
\end{align}
where $h(x,t)$ is the interface height, $\nu$ the diffusivity, and $\lambda _0$ the strength of the slope dependent growth
(with $\lambda_0=0$ giving the EW model). 
$\xi(x,t)$ is the stochastic noise, chosen as a centered Gaussian with short range correlations
\bea
\overline{\xi(x,t) \xi(x',t')}=R_\xi(x-x') \delta(t-t')\,.
\eea
where $\int _x R_{\xi}(x)dx = D$. 
Concerning the space dependence, the most usual choice is to take uncorrelated random disorder, i.e. $R_\xi(x-x') = D \delta(x-x')$. 

In one spatial dimension, the KPZ universality class shows a fairly robust anomalous scaling exponent 
\cite{exponent} for the fluctuations of the height of the interface $h(x,t) \sim t^{1/3}$
and indeed such anomalous behavior at large time has been proved for 
several discrete solvable models  \cite{Johansson2000,spohn2000,png,ferrari1}  which are believed 
to belong to the KPZ class. 
However, this exponent is only one of the facets of the universality of the KPZ equation: 
further universal information is encoded in the full probability distribution function (PDF) of these fluctuations, 
but their exact calculation is an extremely difficult task which is complicated by the fact that, even after a long time, 
the system keeps some memory of the initial conditions. 
Remarkably, these initial conditions can be classified in a few subclasses, each leading to a distinct universal result for the 
statistics of the height field at large time \cite{reviewCorwin,spohnreview}. 
Impressive theoretical progress has been recently achieved and  has led to exact solutions 
directly on the continuum KPZ equation for the wedge (or droplet)
\cite{spohnKPZEdge,we,dotsenko,corwinDP}, flat \cite{we-flat,we-flatlong} and stationary \cite{SasamotoStationary} initial conditions. In the first two cases the PDF of the height $h(x,t)$ at a given point converges at large time to the so-called Tracy-Widom 
GUE (Gaussian unitary ensemble) and GOE (Gaussian orthogonal ensemble) universal distributions \cite{TW1994}, 
 for droplet and flat initial conditions respectively. 
Further impetus to the  field has been given by recent experiments on turbulent liquid crystals \cite{exp4,exp5}
in which these two long-time predictions have been confirmed with high accuracy. 

In the literature, much emphasis has been given to the long time limit, mainly because of the connection with 
random matrix theory valid for all models belonging to the KPZ class. {However, in the general case the scale of the fluctuations heavily depends on the microscopic details of the model: for instance the exact value of
  the mean and the variance of the height fluctuations is known  only for few solvable discrete models. On the contrary, the limit of high diffusivity/weak noise allows the complete determination of the scale of the fluctuations
  as a function of only three parameters $\lambda_0$, $\nu$ and $D$. All the other microscopic details such as the disorder correlations or the lattice effects are relevant only at very short time, $t <t_f$. In particular  the above mentioned exact solutions for the KPZ height distributions are valid for arbitrary times $t>t_f$ in the limit of high diffusivity.}
  
Indeed these solutions can be expressed in terms of Fredholm determinants with rather complicated kernels, 
from which it is not always easy to extract the limiting behavior for long and short times. 
It is then interesting to obtain, by simpler means, the small time behavior in an explicit form, and to confirm it in numerical simulations. 
This has been achieved in the case of the droplet initial conditions \cite{we,ProlhacNumerics}, 
and the aim of this paper is to present a similar result in the case of the flat initial condition. 
As discussed in more details below, there are generically three time regimes: 
\begin{itemize}
\item[(i)] a non-universal very short time regime $t \sim t_f$ where the growth depends on the short scale details of the system 
(e.g. small deviations from the flat initial condition, the precise form of $R_\xi(x)$, etc\dots); 
\item[(ii)] a short time regime $t_f \ll t \ll t^*$ where the crossover from the EW to the KPZ regime takes place; 
\item[(iii)] a large time regime $t \gg t^*$ where KPZ scaling holds. 
\end{itemize}
{In the high diffusivity limit, since $t^*$ is fixed to be very large,
 the height distribution can be exactly computed for all times $t\gg t_f$.
Conversely, in the low diffusivity limit, the  height distribution
depends on the microscopic details of the system for all times and, 
when  $t\to \infty$,  these non-universal details affect only the typical 
scale of the height fluctuations which are Tracy-Widom distributed.}

The paper is organized as follows. In the next section we discuss the 
mapping of the KPZ equation to the directed polymer. 
In Sec. \ref{momsec}, we report the small time expansion of the moments of the partition function of the DP obtained in 
Ref. \cite{we-flatlong}
and we check them in numerical simulations. In Sec. \ref{seccum} we calculate analytically the small time 
expansion of the connected moments of the height field and in Sec. \ref{numerics} we check them by numerical simulations. 
Three appendices contain some more technical calculations.

\section{Mapping to the directed polymer}

Via the Cole-Hopf transformation, the KPZ equation (\ref{kpzeq}) can be mapped onto the directed polymer in a 
random environment which is an equilibrium  statistical physics problem \cite{KPZ,directedpoly,Burgers}. 
A growth starting from a droplet initial condition is mapped onto a fixed endpoints polymer, 
while a flat initial surface translates to a directed polymer with one endpoint fixed and the other free \cite{we-flatlong}. 
Indeed the canonical partition function of a directed polymer $x(\tau)$ at temperature $T$ in a random environment is 
defined in the continuum by the path integral
\be  \label{Zxy}
Z(x,t|y,0)=\int _{x(0)=y} ^{x(t)=x} \!\!Dx e^{-\frac{1}{T} \int _0 ^t d\tau [\frac{1}{2}(\frac{dx}{d\tau})^2+V(x(\tau),\tau)]}\,,
\ee
and maps to the KPZ equation after the identifications
\bea
\frac{\lambda _0}{2 \nu} h  = \ln Z, \qquad 2 \nu = T , \qquad  \lambda_0 \xi(x,t) = - V(x,t)\,.
\eea 
A Gaussian noise $\xi(x,t)$ corresponds to a random potential $V(x,t)$ which is a centered Gaussian with correlator 
$\overline{V(x,t)V(x',t)}=R_V(x-x') \delta(t-t')$ with $R_V(x)=\lambda_0^2 R_\xi(x)$. 
The white noise in KPZ equation corresponds in polymer language to disorder with $\delta$-correlations
\bea
\overline{V(x,t)V(x',t)}=\overline{c}\delta(t-t')\delta(x-x')   , \qquad \bar c = D \lambda_0^2 \,.
\eea
This mapping is valid in the bulk and does not depend on the KPZ initial condition which translates into conditions for the 
endpoints of the polymer. 
For the KPZ equation with flat initial condition, one should consider 
the partition sum with one fixed endpoint (at $x$) and another free (at $y$) \cite{we-flat,we-flatlong}
resulting in the partition function
\bea
Z(x,t) = \int_{-\infty}^{\infty} dy Z(x,t |y, 0)\,.
\label{Zint}
\eea

The recent analytical progress has been made possible by the calculation of  the moments $\overline{Z(x,t)^n}$ 
of the DP partition sum. By replicating the partition function $Z(x,t)$, the DP is mapped \cite{kardareplica} 
onto the quantum mechanics of a bosonic system of $n$ particles interacting with an attractive delta-function potential, 
i.e. the celebrated Lieb-Liniger model \cite{ll}. 
This model is integrable via the Bethe Ansatz and the eigenstates are known for both repulsive \cite{ll} and attractive interactions 
\cite{m-65} which is the case of our interest. 
The moments can be expressed as a sum over these eigenstates \cite{we-flat,we-flatlong} 
(generically labeled by $\mu$ in the following)
\begin{multline}
\overline{Z(x,t)^n}=\\ \sum _\mu \frac{\Psi _\mu ^{*}(x,\cdots ,x)}{||\mu ||^2} e^{-t E_{\mu}} \int _{-\infty} ^{\infty} \prod _{j=1} ^{n} dy_j  \Psi _{\mu} (y_1,\cdots,y_n),
\label{Znint}
\end{multline}
in terms of the many-body wave-function $\Psi _{\mu} (y_1,\cdots,y_n)$ and of the eigenenergies $E_\mu$ of the state $\mu$. 
In the infinite system the eigenstates are easily enumerated, being organized in clusters of bound particles, called strings. The norms of the states $||\mu ||$ and the equal points wave functions have simple expressions \cite{cc-07} and lead to the 
time-dependent PDF starting from a droplet initial condition \cite{we,dotsenko}. 
The integral over the $y_i$ in Eq. (\ref{Znint}), necessary to treat the flat initial condition, is more delicate but 
was handled in Refs. \cite{we-flat,we-flatlong} leading to 
the moments $\overline{Z(x,t)^n}$ for arbitrary $n$. From these 
the moment's generating function at all times has been written in terms of  a Fredholm Pfaffian \cite{we-flat,we-flatlong} 
(the square root of a Fredholm determinant). 
This allowed to prove that the PDF of $\ln Z(x,t)$, i.e. of the height field $h(x,t)$, 
converges at large times to the GOE Tracy-Widom distribution.

Here we follow a different route. We recall in the next section the exact expressions for the lowest moments $n=2,3,4$
and from them we extract the small time cumulants of $\ln Z$, i.e. of the KPZ height field.

\section{Moments $\overline{Z^n}$ and their small time behavior} 
\label{momsec}

For flat initial condition, the one point distribution of $Z(x,t)$ does not depend on $x$ because of translational invariance.
Thus  in the following,  we simply denote 
\bea
Z \equiv Z(x,t) \,.
\eea 
Since we are dealing with an initially flat interface we must have $\overline{Z^n}=1$ at $t=0$ (which is a non-trivial 
condition in terms of the Bethe Ansatz). 
Taking the average of Eq. (\ref{Zxy}) over the Gaussian disorder gives the mean partition function
\bea
\overline{Z(x,t | y,0)} = \frac{1}{\sqrt{2 \pi T t}} e^{- \frac{(x-y)^2}{2 T t}} e^{\frac{R_V(0)}{2 T^2} t } \,,
\eea
and so from the integral in Eq. (\ref{Zint}) we have 
\be 
\overline{Z}=e^{v_0 t}, \quad {\rm with}\;\; v_0=\frac{R_V(0)}{2 T^2}.
\ee 
To eliminate this non-universal (self-energy) contribution, it is convenient to define
\bea
z = Z/\overline{Z}\,,
\eea 
which by construction satisfies $\overline{z}=1$ at all times. (Note that in Ref. \cite{we-flatlong} the self-energy contribution 
was omitted, but we indicate it here explicitly for later purposes. What is called $Z$ in Ref. \cite{we-flatlong} is thus $z$ here.) 
This will be useful later for comparison with the numerical simulation of lattice models.

All results for the continuum DP and KPZ models are expressed in terms of a dimensionless parameter
\be
 \lambda = \Big(\frac{t}{4 t^*}\Big)^{1/3}\,, \quad {\rm with}\;\;
 t^* = \frac{2 T^5}{\bar c^2} = \frac{2 (2 \nu)^5}{D^2 \lambda_0^4} \,,
 \label{lambdadef}
\ee
where, in the language of the DP, $t^*$ is the crossover time scale between the Brownian diffusion at small time (i.e. $\lambda < 1$) 
and the glassy large time behavior (i.e. $\lambda > 1$). 
Within the  context of  the growth model, $t^*$ is the crossover scale between  Edwards-Wilkinson and KPZ regimes. 
Note that a spatial crossover scale can be also defined as $x^* = \sqrt{ \nu t^*} = T^3/\bar c=(2 \nu)^3/D^2 \lambda_0^2$. 
Both scales become very large in the large diffusivity limit or, equivalently, in the weak noise limit. 

It is important to recall that for any given microscopical model with a cutoff (e.g. a lattice model) there are additional time/space scales. 
The easiest example  is the same continuum KPZ equation (or DP model) with a disorder correlated over a 
non-zero correlation length $r_f$, i.e. $R_\xi(x)$ is a function decaying on a scale $r_f$. 
Then, it is easily shown \cite{we} that if $x^* \gg r_f$ one can replace
$R_\xi(x) \to D \delta(x)$ in which $D=\int dx R_\xi(x)$. 
More generally, the condition for the existence of
the universal short time regime studied here is that $t^*$ and $x^*$ must be much larger than any characteristic microscopic
scale --generically called $r_f$ and $t_f$ here-- such as the lattice spacing for a lattice model. Note also that
if the initial condition is not perfectly flat on scales of the order of $r_f$, this will also not affect any result as long as $x^* \gg r_f$. 
Of course, the very short time/space regime with $t\leq t_f$ and $x\leq r_f$ is non-universal. 

We now recall the results of Ref. \cite{we-flatlong} for the four lowest moments, together with their small time (i.e. small $\lambda$) behavior
\begin{eqnarray}
\overline{z^2} & =& e^{2 \lambda^3} [1+ {\rm erf}(\lambda^{3/2} \sqrt{2})]  \label {eqz2} \\
&=&1+2\sqrt{\frac{2}{\pi}} \lambda ^{3/2}+2\lambda ^3 + \frac{8}{3} \sqrt{\frac{2}{\pi}} \lambda^{9/2} + O(\lambda ^{6})\,, \nonumber
\end{eqnarray}
with (note the misprint in Ref. \cite{we-flatlong} for
the definition of the error function) ${\rm erf}(z)=\frac{2}{\sqrt{\pi}} \int_0^z dt e^{-t^2}$, 
\begin{eqnarray}
\overline{z^3} &=&  4 e^{8 \lambda^3 } - 2 e^{2 \lambda^3 }  -  2 e^{8 \lambda^3 } {\rm erfc}(\lambda^{3/2} 2 \sqrt{2})  \label{eqz3}
 \\&&+e^{2 \lambda^3  } {\rm erfc}(\lambda^{3/2} \sqrt{2})  \nonumber\\
& =&1+6\sqrt{\frac{2}{\pi}}\lambda ^{3/2} +14 \lambda ^3+ 40
   \sqrt{\frac{2}{\pi }} \lambda
   ^{9/2}+O(\lambda^{11/2})\,,  \nonumber
\end{eqnarray}
with ${\rm erfc}(x)=1-{\rm erf}(x)$, and finally  
\begin{eqnarray}
 \overline{z^4} &=&   8 e^{20 \lambda^3} - 8 e^{8 \lambda^3} - 4  e^{20 \lambda^3} {\rm erfc}(3 \sqrt{2} \lambda^{3/2})\\&&
-4  e^{8 \lambda^3}  [-2 {\rm erfc}(2 \lambda^{3/2}) + e^{12 \lambda^3} {\rm erfc}(4 \lambda^{3/2})]
  \nonumber\\
&& +  48  \int_0^\infty dx 
\frac{\left(\frac{2 x+1}{\sqrt{4 x+1}}-\frac{\sqrt{2 x+1}}{x+1}\right) e^{-8 \lambda^3 x}}{4 \pi  (4 x (x+3)+5)}   \nonumber\\
& =& 1 + 12 \sqrt{\frac{2}{\pi}} \lambda^{3/2}  + \Big(44+\frac{24}{\pi}\Big) \lambda^3 \nonumber\\&&
+  8 (21 + 8 \sqrt{2})  \sqrt{\frac{2}{\pi}} \lambda ^{9/2} + O(\lambda ^{6}) \,,
\end{eqnarray}
where the series expansion of the integral is performed in Appendix \ref{AA}.

\begin {figure}
\includegraphics[width=0.48 \textwidth]{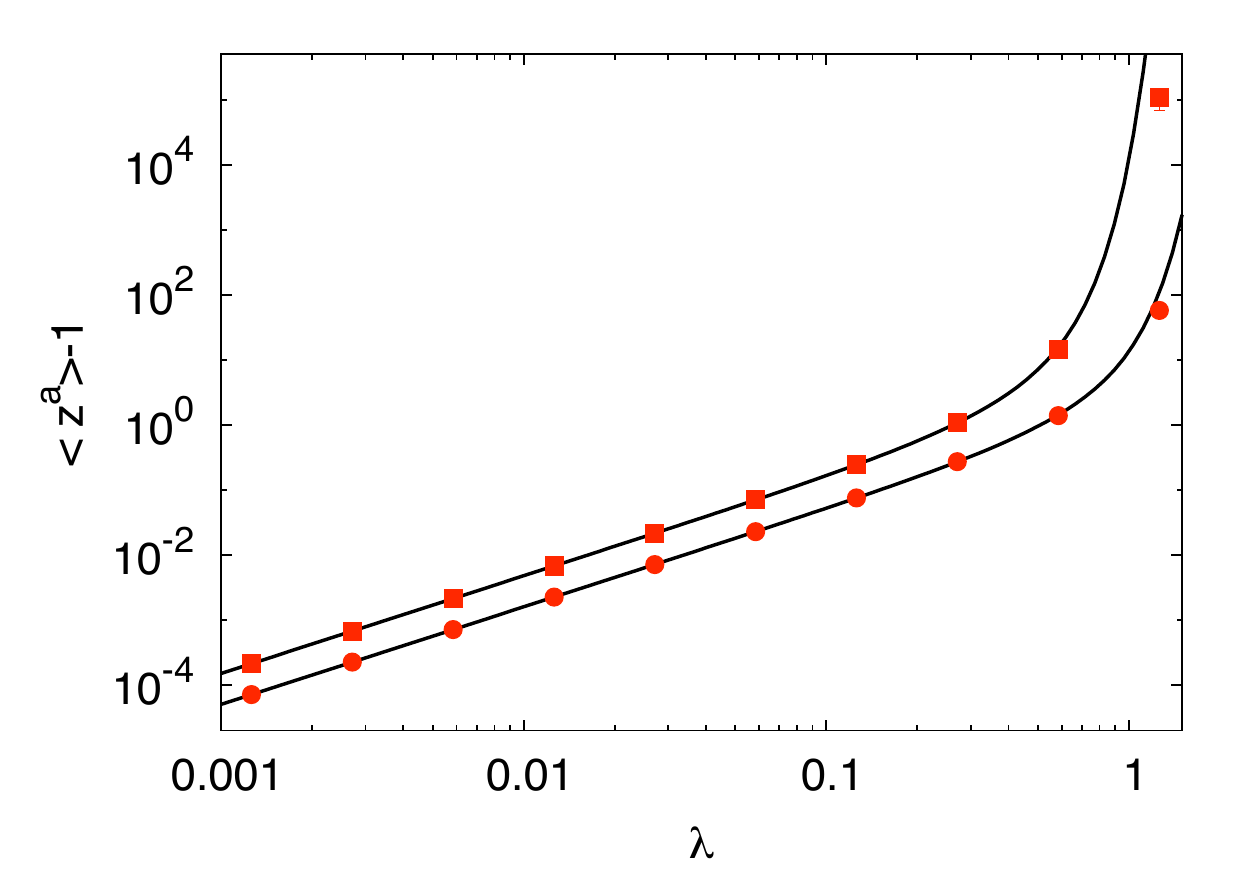}
\caption{(Color online) From top to bottom the moments $\overline{z^3}-1$ ({solid line, red squares}) and $\overline{z^2}-1$ 
(solid line, red circles) 
for different values of $\lambda$. 
Solid lines correspond to the analytical predictions in Eqs. (\ref{eqz3}) and  (\ref{eqz2}). 
Averages are performed over $15 \cdot 10^6$ samples of size $t=512$ and $\overline{c}=1$. There are no adjustable parameters.} 
\label{fig1}
\end {figure} 

Before embarking in the calculation of the cumulants of $\ln Z$, we now report the results of numerical simulations for the 
determination of $\overline{z^n}$ for $n=2,3$. 
As explained in more details in Sec. \ref{numerics}, the simulations are performed for a directed polymer on a square lattice. 
We also consider the high temperature limit which ensures that all details of the lattice become irrelevant and 
the results can be expressed as functions of the single parameter $\lambda$. 
The procedure and the identification of $\lambda$ on the lattice have been introduced already 
in \cite{we,highT} and are described again in Sec. \ref{numerics}.
We report the numerical data for $\overline{z^n}$ in  Fig. \ref{fig1} which are found to be in excellent agreement with  
our analytic  predictions up to $\lambda \approx 0.6$, while some deviations at larger $\lambda$ are evident.
These deviations are caused by the undersampling due to the growing importance of the tails in the distribution of $z$, 
and will be properly explained in Sec. \ref{numerics}.

\section{cumulants of $\ln Z$ at small time}
\label{seccum}

From the above formulas for $\overline{z^n}$ and following the procedure described in Appendix \ref{B}, 
we obtain the small $\lambda$ (i.e. small time) expansion of the first four cumulants of the free energy 
\bea \label{log1}
\overline{\ln z}&=& -\sqrt{\frac{2}{\pi }} \lambda
   ^{3/2}+\Big(\frac{5}{3}-\frac{6}{\pi }\Big)
   \lambda ^3 \\&& + \Big(\frac{106}{3}-16 \sqrt{2}-\frac{40}{\pi } \Big) \sqrt{\frac{2}{\pi }} \lambda^{9/2} 
   +O(\lambda^6) ,\nonumber \\
 \label{log2}
\overline{(\ln z)^2}^c&=&
2 \sqrt{\frac{2}{\pi }} \lambda^{3/2}+\Big(\frac{20}{\pi }-6\Big) \lambda^3 \\ &&
 + \Big(\frac{176\sqrt{2}}{3}+\frac{512}{3 \pi } -\frac{412}{3}\Big) \sqrt{\frac{2}{\pi }} \lambda^{9/2} +O(\lambda^6) , \nonumber
\\
  \label{log3}
\overline{(\ln z)^3}^c&=& \frac{8 (\pi -3) \lambda ^3}{\pi } \\ &&
+ \Big(248-96 \sqrt{2}-\frac{352}{\pi }\Big) \sqrt{\frac{2}{\pi }} \lambda^{9/2}
+O(\lambda^6),\nonumber  \\
\overline{(\ln z)^4}^c&=& \Big[64 \sqrt{2}+\frac{320}{\pi } -192 \Big] \sqrt{\frac{2}{\pi }} \lambda^{9/2}+O(\lambda^6), \label{log4}
\eea
and of course $\overline{(\ln Z)^p}^c=\overline{(\ln z)^p}^c$ for $p \geq 2$.
As explained in  Appendix \ref{B}, in order to compute the next term $O(\lambda^6)$ in the small time expansion, 
or the fifth and higher cumulants,  we would need the fifth moment $\overline{z^5}$ that we did not analyze here, 
but which is in principle known \cite{we-flatlong}.
A simple check can be performed on these formulas, namely one can compute the series expansion 
in $\lambda^{3/2}$ of $\exp( \sum_{p=1}^4 \frac{n^p}{p!} \overline{(\ln z)^p}^c )$ and
check that the expansion of all $\overline{z^n}$ for $n=1,2,3,4$ given above is reproduced up to 
order $o(\lambda^{9/2})$. Although this procedure also allows to derive Eqs. (\ref{log1}-\ref{log4}) 
by adjusting the coefficients of the series expansion in $\lambda^{3/2}$, 
the method described in Appendix \ref{B} is more systematic.

For short times, the dominant term in the PDF is the variance $\overline{(\ln z)^2}^c$ which increases as
$t^{1/2}$. Using $\ln Z= \lambda_0 h/(2 \nu)$, one finds 
\bea
\label{dominantKPZ}
\overline{h^2}^c=D\,\sqrt{\frac{t}{2\pi \nu}}+O(\lambda_0 ^2 t)\,.
\eea
Hence the first term of the expansion of $\overline{h^2}^c$ is independent of $\lambda_0$, 
the coefficient of the non-linear growth in the KPZ equation. 
It corresponds to the Edwards-Wilkinson Gaussian scaling regime $\delta h \sim t^{1/4}$, 
also found in Appendix \ref{C}, cf.  Eq. (\ref{ewperturbative}), where we derive the leading short time behavior 
for the average height  and variance using perturbation theory directly on the KPZ equation.

The third and fourth cumulants behave as $t$ and $t^{3/2}$ respectively, suggesting that the
fourth cumulant is subdominant and that the first corrections to the EW gaussian scaling 
are given by the third cumulant as $\delta h \sim t^{1/3}$, which is the form of the KPZ scaling.
Indeed, it is interesting that the third cumulant is linear in $t$ {\it both at short and large times} 
(but with an amplitude going from $\frac{8(\pi-3)}{\pi}=0.360563$ to $\mu _{3}^{GOE}=0.598268$, see below). 

From the above we can compute the skewness 
\bea
 \gamma_1 &=& \frac{\overline{(\ln Z)^3}^c}{[\overline{(\ln Z)^2}^c]^{3/2}} = \frac{2^{3/4} (\pi -3)}{\sqrt[4]{\pi }} \lambda^{3/4}
\\ && +\frac{\pi   \left(67-48 \sqrt{2}+9 \pi \right)-86}{(2 \pi )^{3/4}}  \lambda ^{9/4}+O(\lambda^{15/4}) \nonumber\\
   & =& 0.178865\dots \lambda ^{3/4} + 0.0138\dots \lambda ^{9/4} +O(\lambda^{15/4}),\nonumber
\eea 
which at short times scales as $\gamma_1 \sim t^{1/4}$ and the kurtosis
\begin{multline}
 \gamma_2 = \frac{\overline{(\ln Z)^4}^c}{[\overline{(\ln Z)^2}^c]^2} = (40-24 \pi +8 \sqrt{2} \pi ) \sqrt{\frac{2}{\pi}} \lambda^{3/2} + O(\lambda^3)\\
= 0.115565\dots  \lambda^{3/2} + O(\lambda^3) \,,
\end{multline}
which scales as $t^{1/2}$.

\begin {figure}
\begin{center}
\includegraphics[width=0.48 \textwidth]{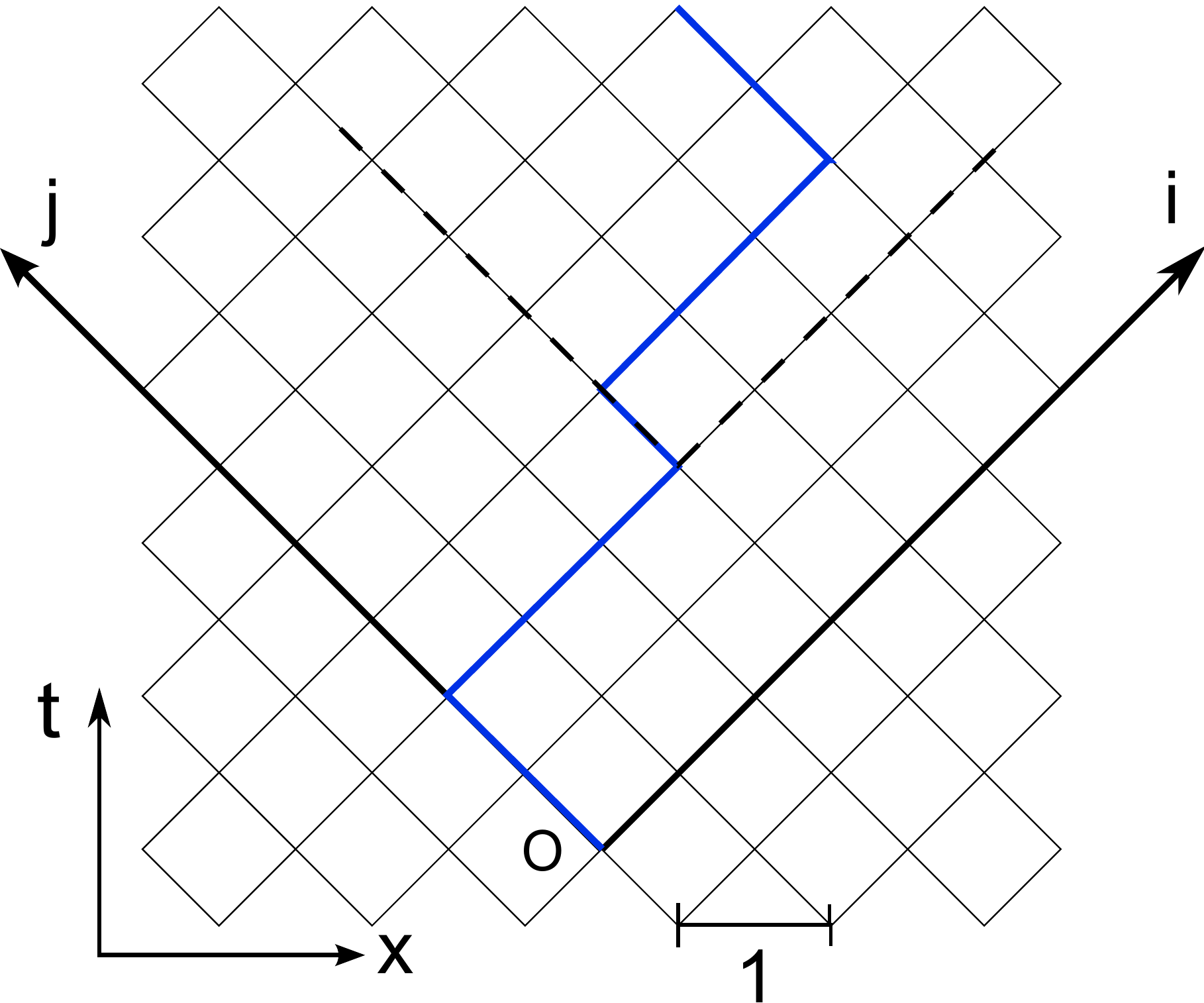}
\end{center}
\caption{(Color online) Sketch of the directed polymer model analyzed in numerical simulations. 
The blue solid line corresponds to a polymer growing over the square lattice under the hard constraint condition.} 
\label{picture_polymer}
\end {figure}

Now we recall that at large time one can write \cite{we-flat,we-flatlong}
\bea
\frac{\lambda_0 h}{2 \nu} = \ln Z = v_{\infty} t + \lambda \eta_t ,
\eea 
such that $\eta_t$ converges to the GOE Tracy-Widom distribution $\displaystyle \lim_{t \to \infty} {\rm Prob}(\eta_t < s) = F_1(s)$. 
The skewness and kurtosis thus converge for large times to their GOE values
\bea
&& \gamma_1 \to \gamma_1^{GOE} = 0.29346452408\dots, \\
&& \gamma_2 \to \gamma_2^{GOE} = 0.1652429384\dots,
\eea 
consistent with a crossover for $\lambda \approx 1.6 \pm 0.3$. 
The amplitude of the (non-fluctuating) linear term is non universal, $v_{\infty} = v_0 - \bar c^2/12 = v_0 - 
 D^2 \lambda_0^4/12$, where $v_0=\frac{R_V(0)}{2 T^2}=\frac{\lambda_0^2 R_\xi(0)}{8 \nu^2}$ 
is the amplitude at short time (after the very short time regime $t \gg t_f$). Note
that the difference $v_{\infty}-v_0$ is universal. At large time one also has that
$\overline{\ln z} = \overline{\ln Z} - \ln \overline Z=  \lambda \mu_1 -  D^2 \lambda_0^4 t/12$ is universal,
where $\mu_1^{GOE}=-1.2065335745820\dots$ is the mean of the TW distribution, while
$\overline{(\ln Z)}^c \to \lambda^2 \mu_2$ where $\mu_2^{GOE} =1.607781034581\dots$ 
is the variance of the TW distribution.

\section{Numerical results}\label{numerics}

\begin {figure}
\begin{center}
\includegraphics[width=0.48 \textwidth]{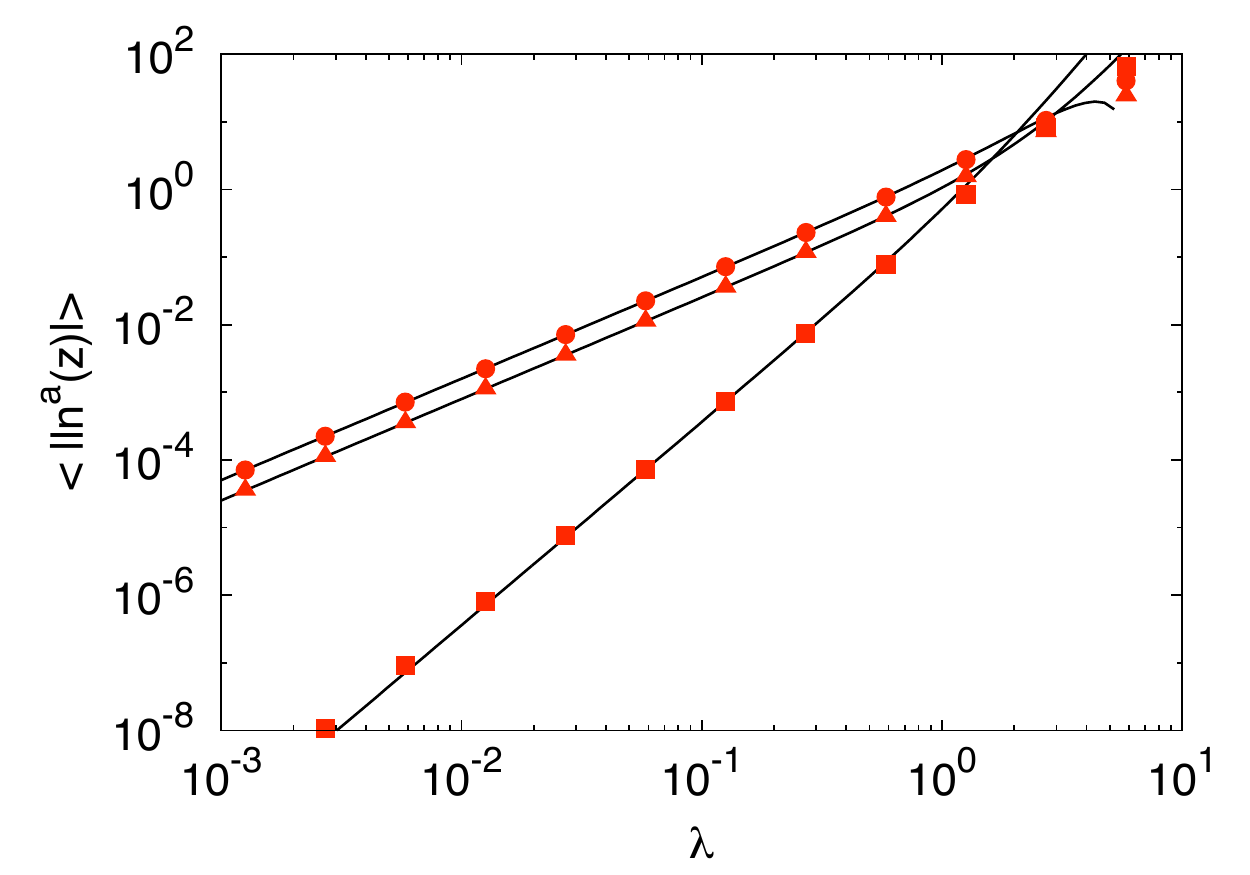}
\end{center}
\caption{(Color online) From top to bottom, the cumulants ($15 \cdot 10^6$ samples) $\overline{\ln ^2 z}^c$ (solid line, red circles),  $-\overline{\ln z}^c$ 
(solid line, red triangles) and $\overline{\ln ^3 z}^c$ (solid line, red squares) for $t=512$. The solid lines  are the analytical predictions 
in Eqs. \eqref{log1}, \eqref{log2}, and \eqref{log3} up to $O(\lambda^{9/2})$, with $\overline{c}=1$. There are no adjustable parameters.} 
\label{cumulants}
\end {figure} 

\begin {figure}
\begin{center}
\includegraphics[width=0.48 \textwidth]{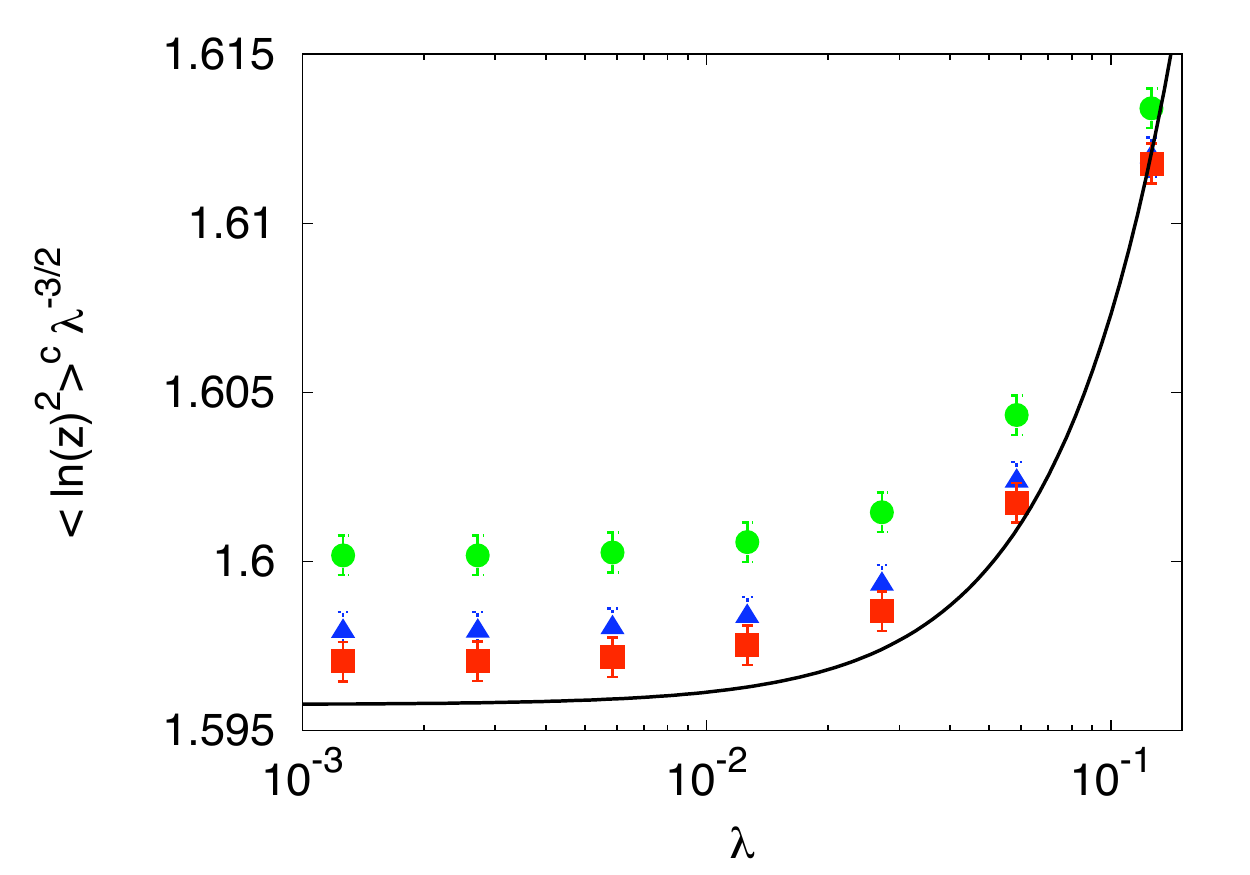}
\end{center}
\caption{(Color online) Finite-size effects for $\overline{\ln ^2 z}^c$. Solid line: analytical prediction Eq. \eqref{log2}. Numerical data, from top to bottom $t=128$ (green circles), $t=256$ (blue triangles), $t=512$ (red squares). Averages are performed over $15 \cdot 10^6$ samples with $\overline{c}=1$.} 
\label{finite_size}
\end {figure} 

\begin {figure}[b]
\begin{center}
\includegraphics[width=0.48 \textwidth]{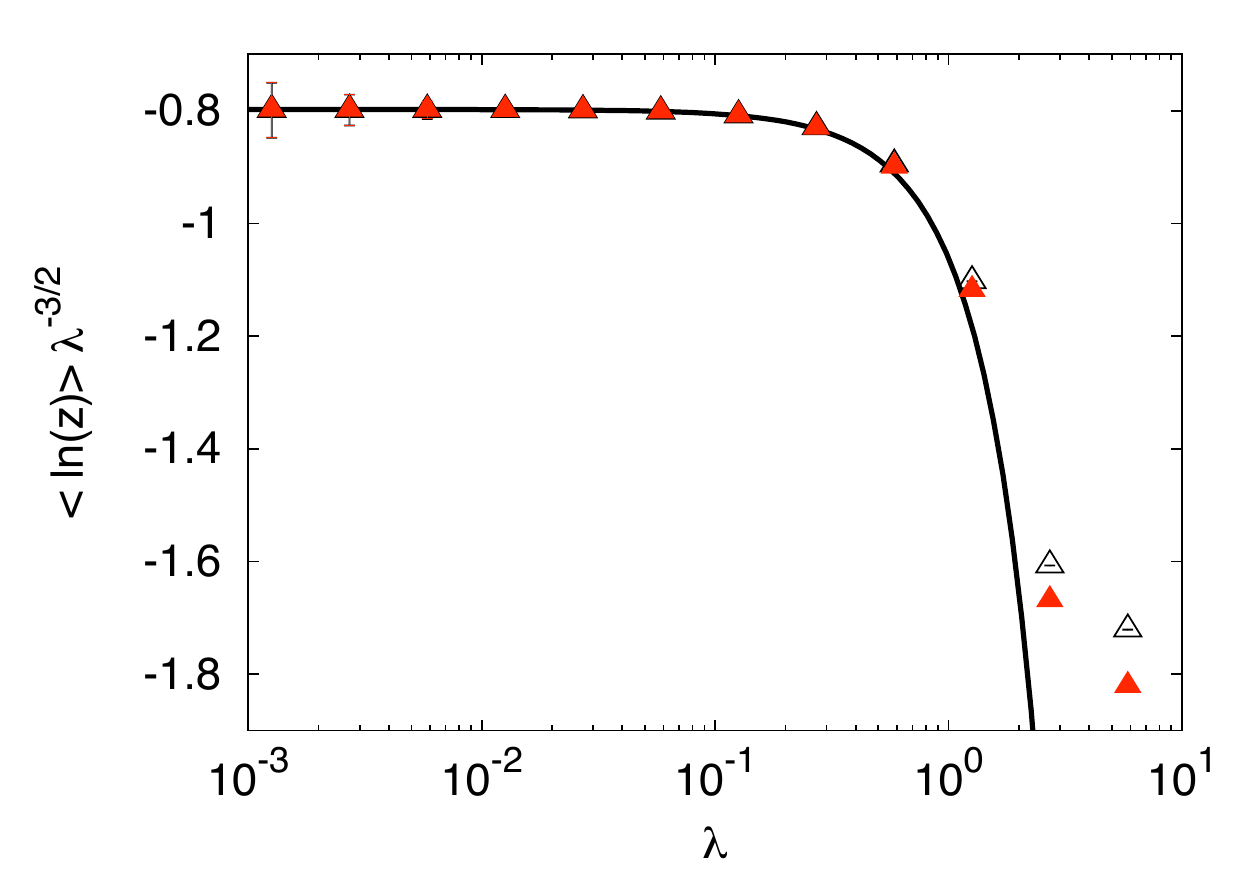}
\end{center}
\caption{(Color online). Finite-size effects for $\overline{\ln z}^c$. Solid line: analytical prediction Eq. \eqref{log1}. Numerical data, from top to bottom $t=256$ (empty triangles), $t=512$ (full triangles). Averages are performed over $15 \cdot 10^6$ samples with $\overline{c}=1$. The large error bars when $\lambda < 10^{-2}$ are due to the vanishing value of $\overline{\ln z}^c$ as $\lambda \rightarrow 0$.} 
\label{finite_size2}
\end {figure}

Numerical simulations are performed for the square lattice model depicted in Fig. \ref{picture_polymer}. 
Directed paths grow along the diagonals of the lattice with only $(0,1)$ or $(1,0)$ moves (hard constraint condition), 
starting in $(0,0)$ and with the second end left free. 
To each site of the lattice is associated an independent and identically distributed random number $V(x,t)$ (here we use  a Gaussian distribution with variance equal to $1$). 
The time coordinate is given by $t=i+j$ and the space coordinate by $x=(i-j)/{2}$ (see Fig. \ref{picture_polymer}).
The partition sum over all paths $\gamma_t$ growing from $(0,0)$ up to time $t$ is defined as
\bea
Z(t)=\sum _{\gamma_t} \exp\Big[{- \beta\!\! \sum _{(x,\tau) \in \gamma_t} V(x,\tau)}\Big]\,.
\eea
The partition function satisfies the following transfer matrix recurrence relation implemented in our simulation
\bea
Z_{x,t+1}=(Z_{x-\frac{1}{2},t}+Z_{x+ \frac{1}{2},t}) e^{-\beta V_{x,t+1}}\,,
\eea
with $Z_{x,0}=\delta_{x,0}$. The free end partition function is computed by summing over all endpoints $Z(t)=\sum _x Z(x,t)$.  
To avoid numerical instabilities we divided all partition functions at fixed $\tau$ by the biggest one and record its logarithmic value. 
As in the model in the continuum, also on the lattice $Z(x,t)$ grows exponentially in time, as can be 
seen by averaging the sum over all possible paths
\bea
\overline{Z(t)}=\sum_{\gamma_t} \prod_{x \in \gamma_t} \overline{e^{- \beta V(x)}} = 2^t e^{\beta ^2 t/2}\,.
\eea 
For this reason we work numerically with the ratio $\ln(z)=\ln(Z/\overline{Z})$, 
which remains small, but exhibits strong fluctuations.

In the limit of high $T$, the statistical fluctuations of $z$ only depend on the unique dimensionless variable
\bea \label{lambdadiscrete}
\lambda = \Big( \frac{\overline{c}^2 \kappa t }{8 T^5} \Big) ^{1/3},
\eea
which is the lattice version of Eq. (\ref{lambdadef}). 
Note that the scaling $T \rightarrow T/\kappa$, $\overline{c} \rightarrow \overline{c} /\kappa^2$ allows to go from the discrete model variables Eq. (\ref{lambdadiscrete}) to the continuous model variables Eq. (\ref{lambdadef}).

In the high temperature regime, the parameters $\kappa$ and $\overline{c}$ can be computed explicitly  \cite{highT}. 
Indeed $\overline{c}^2$ is just the variance of the uncorrelated random numbers.
Instead $\kappa$ can be extracted from the model without disorder, for which the polymer behaves like a particle diffusing on a 
one-dimensional lattice ($x$ being the particle position at time $t$). 
The mean square displacement of the particle is given by $\langle x^2 (t) \rangle _T = T t/\kappa$. 
Within the normalization used in this paper, we have $\kappa = 4 T$. 

Using this algorithm, we have numerically computed the cumulants on the square lattice at high temperature and
compared them with the analytic predictions in Eqs. \eqref{log1}, \eqref{log2}, and \eqref{log3}. 
The data for $\overline{\ln ^n z}^c$ for $n=1,2,3$ are reported in Fig. \ref{cumulants}.
The agreement with the analytical predictions is excellent, which is even more impressive when we 
consider that these figures are produced {\it without any fit parameter}. 
In Figs. \ref{finite_size} and \ref{finite_size2} we show in more details the convergence to the theoretical 
value for a fixed value of $\lambda$ as a function of polymer length. 
The increase of the polymer length $t$ is equivalent to heating up the system, 
hence approaching the universal prediction of the high temperature regime.

\begin {figure}
\begin{center}
\includegraphics[width=0.48 \textwidth]{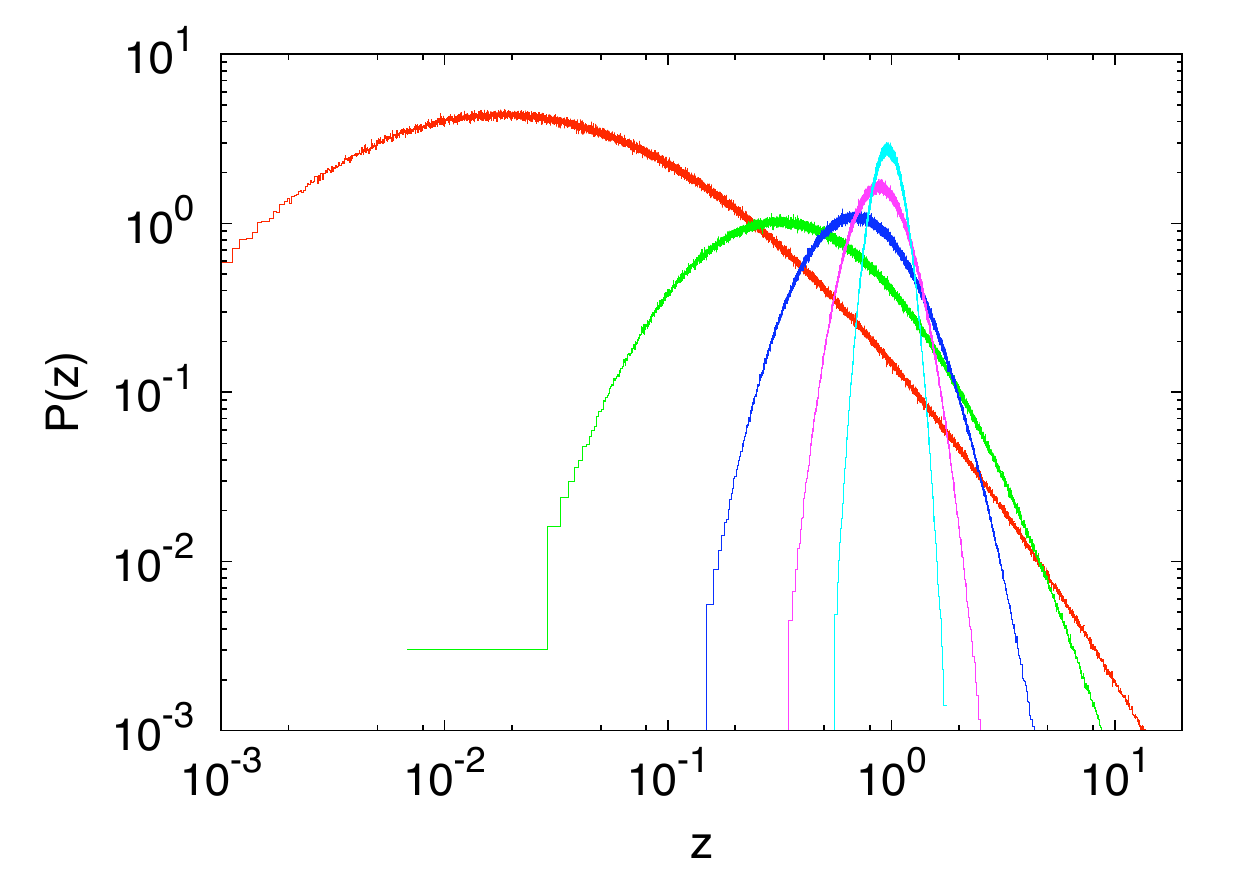}
\end{center}
\caption{(Color online) ($P(z)$ with $z=\frac{Z}{\overline{Z}}$ for $\lambda = 1.26 , 0.58, 0.27 , 0.126$ and $0.058$, from left to right. Histograms are obtained from numerical simulations with $t=512$,  $\overline{c}=1$ and $15 \cdot 10^6$ samples. When $\lambda$ is very small, $P(z)$ is self-averaging. When $\lambda$ grows, a heavy tail is developed and $z_{typ}\ll\langle z\rangle = 1$.} 
\label{log_norm}
\end {figure}

\begin {figure}
\includegraphics[width=0.48 \textwidth]{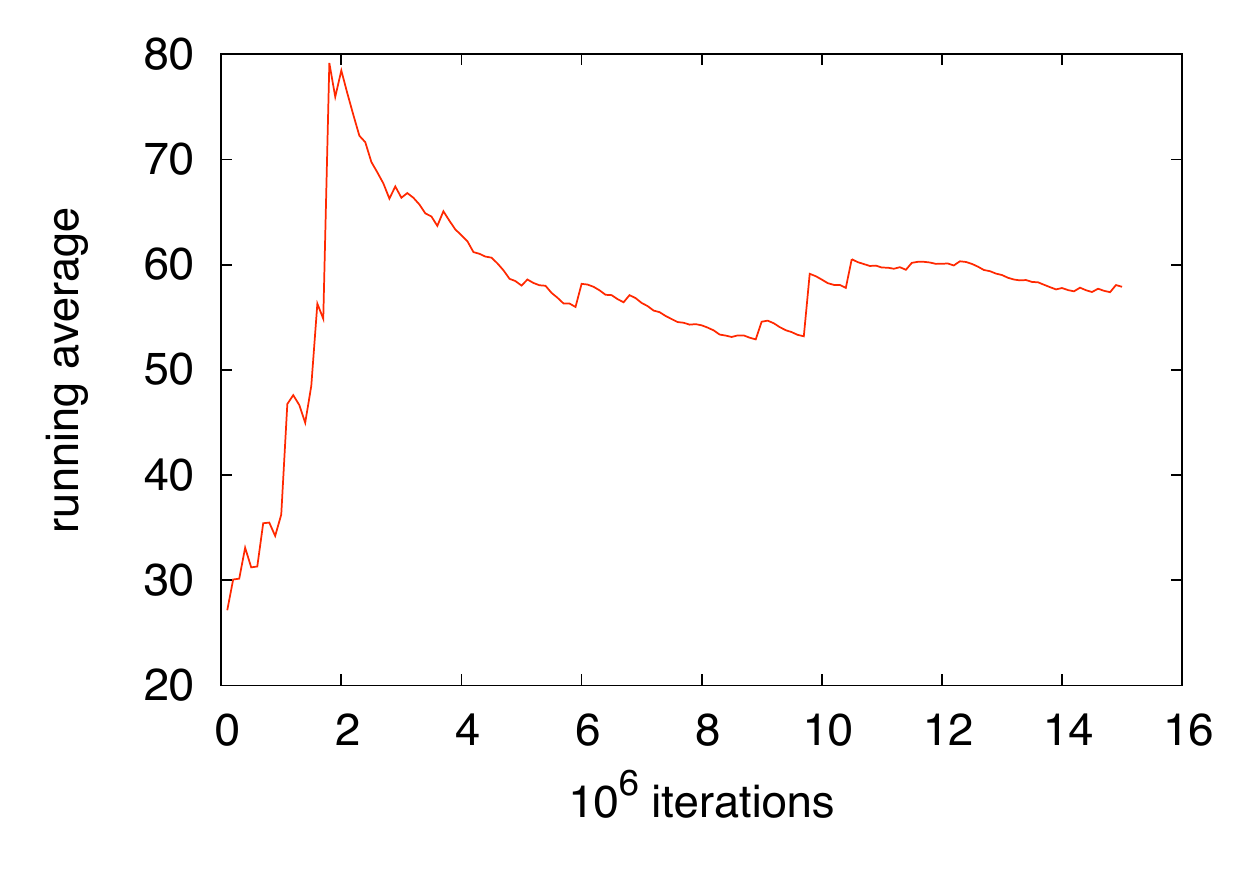}
\caption{(Color online) Estimator of $\overline{z^2}-1$ given by $M_N=\frac{1}{N} \sum _{i=1}^{N} z_i^2 -1$, where $z_i$ is the rescaled partition function of a single disorder realization and $N$ is the number of realizations (iterations), for $\lambda=1.26$. The sudden variations around $10 \cdot 10^6$ samples shows that one \textit{single} event contributes to a finite fraction (around $10 \%$) of the whole sum.}
\label{nonaveraging}
\end {figure}

The analytical predictions for the moments of $z$ are exact for all $\lambda$. 
However, we can see in Fig. \ref{fig1} that precision is quickly lost above the threshold $\lambda \sim 1$. 
This is due to the fact that, for large $\lambda$, typical values of $z$ strongly differ from the average value $\overline{z}=1$. 
The moments of $z$ are then dominated by rare occurrences of very large $z$ induced by the presence of heavy tails. 
This is shown in Fig. \ref{log_norm} where we see that, as $\lambda$ grows, the mode of the distribution quickly goes to $0$ 
while the tail becomes fatter. 
A simple example of this peculiar behavior lies in the log-normal probability distribution which is characterized by an 
exponentially small typical value and a heavy tail $\sim e^{-\ln^2 z}$. 
Here, for large $\lambda$, the heavy tail behaves as $e^{-\ln^{3/2} z}$ with the exponent $3/2$ corresponding 
to the Tracy-Widom asymptotic behavior.
In practice, because of this tail, the moments estimators converge extremely slowly, even for important sampling. 
An example is shown in Fig. \ref{nonaveraging}, where we estimate $\overline{z^2}-1$ for $\lambda=1.26$ 
with $N=15 \cdot 10^6$ samples. 
While the prediction from Eq. (\ref{log2}) is $\overline{z^2}-1 =109.023$, we found a value around $60$ from numerical simulations. 
This discrepancy is explained by the central limit theorem which predicts fluctuations of order $\sim (\overline{z^4}/N)^{1/2}$. 
Using Eq. \ref{log4}, we see that $\overline{z^4}$ grows very fast in $\lambda$ and would require 
$N=10^{14}$ samples to have a good estimation of $\overline{z^2}$ for $\lambda=1.26$.

\section{Conclusion}

In this paper we have studied the stochastic KPZ equation with flat initial conditions
and extracted from the results of 
Ref. \cite{we-flatlong} the short time behavior of the connected moments of the distribution of
the height field at a given point. 
In this way,  we have been able to probe universality specifically with 
respect to the introduction of a short scale in the noise correlations, or a discretization in the DP model. 
A wider domain of investigation, 
ranging from step bunching instabilities in crystal growth to ballistic deposition (see e.g. Ref. \cite{spohn2000})
and going beyond the goals of this work is to prove the
universality in a broader sense (including e.g. change in the non linear
KPZ term, as in Ref. \cite{SpohnKrug}, or biased diffusion current, as in Ref. \cite{PimpinellKrug}). 

The importance of the results presented here stems from the proof of the existence of a short time {\it universal} regime
which describes the crossover from the Edward-Wilkinson to the KPZ growth and which
can be observed when the diffusivity is large or the noise is weak. 
We have compared our analytical predictions to high precision numerical simulations of
a discrete model, which shows how this universality arises. 
A part from the  theoretical interest per se, these predictions, valid for all times, should be useful also in future experiments in which 
the parameters of the growth could be varied and controlled in a more refined way so to 
easily access this universal crossover.

\appendix
\section{Expansion of an integral}\label{AA}

We need to compute the small $\lambda$ expansion of the integral
\bea
I = \int_0^\infty dx  f(x) e^{-8 \lambda^3 x},
\eea
where $f(x)$ and its large $x$ expansion are
\begin{multline}
 f(x) = 48 \frac{\left(\frac{2 x+1}{\sqrt{4 x+1}}-\frac{\sqrt{2 x+1}}{x+1}\right)}{4 \pi  (4 x (x+3)+5)}\\  
= \frac{3 }{\pi} x^{-3/2} -\frac{3 \left(21+8 \sqrt{2}\right) }{8 \pi }x^{-5/2}+O(x^{-7/2})\,.
\end{multline}
It is convenient to write
\begin{multline}
I = \int_0^\infty dx  f(x) + \int_0^\infty dx \frac{3}{\pi x^{3/2}
   }  (e^{-8 \lambda^3 x} -1)\\ + \int_0^\infty dx  f_1(x)  (e^{-8 \lambda^3 x} -1) ,
\end{multline}
with $f_1(x) = f(x) -  3 x^{-3/2} /\pi$. Two integrals are easily done, giving
\be 
I = 1 - 12 \sqrt{\frac{2}{\pi}} \lambda^{3/2} + \int_0^\infty dx f_1(x) (e^{-8 \lambda^3 x} -1) ,
\ee
and the remaining integral is $O(\lambda^3)$. 
This can again be written as
\begin{multline}
 \int_0^\infty dx f_1(x) (e^{-8 \lambda^3 x} -1) = (- 8 \lambda^3 ) \int_0^\infty dx x f_1(x) \\
+ \int_0^\infty dx f_1(x)  (e^{-8 \lambda^3 x} -1 + 8 \lambda^3 x) \\
 = \Big(44 + \frac{24}{\pi}\Big) \lambda^3 + \int_0^\infty dx f_1(x)  (e^{-8 \lambda^3 x} -1 + 8 \lambda^3 x),
\end{multline}
where the remaining integral is now $O(\lambda^{9/2})$ and can be split again as
\bea
&& \int_0^\infty dx f_1(x)  (e^{-8 \lambda^3 x} -1 + 8 \lambda^3 x) =  \nonumber \\&& 
= \int_0^\infty dx \Big[-\frac{3 \left(21+8 \sqrt{2}\right)
   }{8 \pi } x^{-5/2}\Big]    (e^{-8 \lambda^3 x} -1 + 8 \lambda^3 x) \nonumber\\
   && + 
\int_0^\infty dx f_2(x)  (e^{-8 \lambda^3 x} -1 + 8 \lambda^3 x), 
\eea
with 
\be
 f_2(x) = f_1(x) + \frac{3 \left(21+8 \sqrt{2}\right)}{8 \pi} x^{-5/2}.
\ee 
Thus, putting together the three pieces, we have 
\begin{multline}
 I = 1 - 12 \sqrt{\frac{2}{\pi}} \lambda^{3/2} + (44 + \frac{24}{\pi}) \lambda^3 \\ -8 (21 + 8 \sqrt{2}) \sqrt{\frac{2}{\pi}} \lambda^{9/2} 
 + O(\lambda^6)\,.
\end{multline}

\section{From the moments of $Z$ to the moments of $\ln Z$} \label{B}

In general the knowledge of the moments $\overline{z^n}$ for some low integer $n$ does not
allow to extract much information about the cumulants $\overline{(\ln z)^n}^c$. In the present case, however,
for small time (small $\lambda$), $z$ is concentrated around its mean value $\overline{z}=1$ and this allows
to obtain the behavior of the cumulants at small times.

Let us write $z=1+u$ with $\overline{u}=0$ and compute its connected moments. 
In order to lighten the notation we introduce the notation $\mu_p \equiv \overline{z^p}^c$.
Using the expressions for  $\overline{z^n}^c$ in the main text,  $\overline{u^n}^c$ are given by
\bea
\mu_2 &=& \overline{u^2} =\overline{z^2}-1 \nonumber\\ &=& 2 \sqrt{\frac{2}{\pi }} \lambda
   ^{3/2}+2 \lambda^3+ \frac{8}{3} \sqrt{\frac{2}{\pi}} \lambda^{9/2} + O(\lambda ^{6}),  \\
\mu_3 &=& \overline{u^3}=(\overline{z^3}-1)-3(\overline{z^2}-1)  \nonumber\\ &=& 8 \lambda ^3+32 \sqrt{\frac{2}{\pi
   }} \lambda ^{9/2}+O(\lambda^6), \\
\mu_4&=& \overline{u^4}^c=  \overline{z^4} - 1 - 4 (\overline{z^3} - 1) + 6 (\overline{z^2} - 1) - 3 (\overline{z^2}-1)^2
      \nonumber\\ &=&  64 \sqrt{2} \sqrt{\frac{2}{\pi }} \lambda^{9/2}+O(\lambda^6).
\eea
Given the above trend it is reasonable to assume that $\overline{z^p}^c = \overline{u^p}^c = O( (\lambda^{3/2})^{p-1})$. 
Based on this assumption,
we want to construct a systematic series expansion of $\overline{(\ln z)^n}^c$ in powers of 
the cumulants of $z$. 
The reasoning is the following. First we write
\begin{multline} \label{ser1}
\sum_{n=1}^\infty \frac{r^n}{n!} \overline{(\ln z)^n}^c
= \ln \overline{z^r} = \ln \overline{(1+u)^r} =\\ \ln\Big(1 + \sum_{k=1}^\infty \frac{r(r-1)..(r-k+1)}{k!} \overline{u^k}\Big) \,.
\end{multline}
Expanding the rhs in powers of $r$, we obtain formally each $\overline{(\ln z)^n}^c$ as an (infinite) series
of the moments $\overline{u^k}$. 
The moments $\overline{u^k}$ can themselves be expressed
as functions of the cumulants $\mu_p$ by writing
\bea \label{ser2}
\overline{e^{w u}} = 1 + \sum_{k=2}^\infty \frac{w^m}{m!} \overline{u^m} = \exp\Big({\sum_{p=2}^\infty \frac{w^p}{p!} \mu_p} \Big),
\eea 
and identifying them order by order in $w$. We can now replace $\mu_p \to a^{p-1} \mu_p$ where $a$ is to be set
to unity at the end. This replacement allows us to keep track of the order in $\lambda^{3/2}$ of each cumulant. 
Using Mathematica, it is now simple to first generate the expansion (\ref{ser2}), truncate it to a given order in $a$, and
then insert it in Eq. (\ref{ser1}). 
During this process, we see  that e.g. $\overline{u^4} = O(a^2)= \overline{u^3}$, i.e. in Eq. (\ref{ser1}) one must keep a few 
more orders compared to Eq. (\ref{ser2}). 
Since we have not computed $\overline{z^5} = O(a^4)$, we can only get our cumulants
up to order $a^3$. Doing so we obtain
\bea
\overline{\ln(z)}&=&
-\frac{a \mu _2}{2}+a^2 \Big(\frac{\mu _3}{3}-\frac{3 \mu _2^2}{4}\Big) \\&& +a^3
   \Big(-\frac{5 \mu _2^3}{2}+2 \mu _3 \mu _2-\frac{\mu _4}{4}\Big)+O(a^4) ,\nonumber\\
 \overline{\ln(z)^2}^c &=& a \mu _2+a^2 \Big(\frac{5 \mu _2^2}{2}-\mu _3\Big)\\&& +a^3 \Big(\frac{32 \mu _2^3}{3}-8 \mu
   _3 \mu _2+\frac{11 \mu _4}{12}\Big)+O(a^4),  \nonumber\\
\overline{\ln(z)^3}^c &=&  a^2 \Big(\mu _3-3 \mu _2^2\Big)\\&&+a^3 \Big(-22 \mu _2^3+15 \mu _3 \mu _2-\frac{3 \mu
   _4}{2}\Big)+O(a^4) ,\nonumber\\
\overline{\ln(z)^4}^c &=& a^3 \Big(20 \mu _2^3-12 \mu _3 \mu _2+\mu _4\Big)+O(a^4).
\eea
Setting $a=1$ and replacing the $\mu_p$ by their actual values above, we find the result given in the text. 

\section{Short time perturbation theory for the KPZ equation}\label{C}

As a further final check, we recover here the leading short time behavior for the first and second cumulants
of the height directly from the perturbative expansion of the KPZ equation. 
We start from the second cumulant which is easier. 
The KPZ equation can be studied in perturbation in $\lambda_0$ which is equivalent to short time. 
This is clear from the definition of $\lambda$ in Eq. (\ref{lambdadef}) which gives the perturbative parameter
$\lambda^{3/2} \propto \sqrt{t/t^*} \propto \sqrt{t} \lambda_0^2 $. 
We can write $h=h^{(0)}+h^{(1)}+\dots$ where $h^{(n)} = O(\lambda_0^n)$. 
With the flat initial condition, the lowest order is just the Edwards-Wilkinson result that in Fourier space is 
\bea
h^{(0)}_{q,t} = \int_0^t dt_1 e^{- \nu q^2 (t-t_1)} \xi_{q,t_1} ,
\eea 
which leads to the variance
\bea
\overline{h^{(0)}_{q,t} h^{(0)}_{q',t}} = 2 \pi \delta(q+q') \int_0^t dt_1 e^{- 2 \nu q^2 t_1} \tilde R_\xi(q) ,
\eea 
where $\tilde R_\xi(q)$ is the Fourier transform of the noise correlator $R(x)$, assumed to be of range $r_f$ in space.  
Then simple algebra gives
\bea
\label{ewperturbative}
\overline{h^{(0)}(x,t)^2} &=& \int \frac{dq}{2\pi} \int_0^t dt_1 e^{- 2 \nu q^2 t_1} \tilde R_\xi(q)  \\
&=& \int \frac{dq}{2\pi} \tilde R_\xi(q) \frac{1 - e^{- 2 \nu q^2 t} }{2 \nu q^2} 
= \frac{D}{\sqrt{\nu}} \sqrt{\frac{t}{2 \pi}} ,\nonumber
\eea 
where the last equation is valid for $t \gg r_f^2/\nu$, i.e. away from the (non-universal) very short time regime. 
Here $D=\tilde{R}_\xi(q=0)=\int dx R(x)$ is the only memory of the short scale details and thus for $t \gg t_f$ 
one can set $\tilde R_\xi(q=0)=D$, i.e. a delta-correlator in space. Using the correspondence $\ln Z = \lambda_0 h/(2 \nu)$ and $\lambda^{3/2} = D \lambda_0^2 (t/8)^{1/2}/(2 \nu)^{5/2}$ one recovers exactly the leading term in Eq. (\ref{dominantKPZ}). 

The discussion of the average height $\overline{h}$ is more subtle because we need to retain $R_\xi(x)$ in an essential way,
as there are non-universal contributions. For a flat initial condition, $\overline{h(x,t)}$ is $x$ independent, 
hence the following equation is exact at all times
\bea
\partial_t \overline{h} = \frac{\lambda_0}{2} \overline{ (\nabla h)^2 } \,.
\eea 
At small time we can use 
\bea
\partial_t \overline{h} = 
\frac{\lambda_0}{2} \overline{ (\nabla h^{(0)})^2 } = \frac{\lambda_0}{4 \nu} \int_q \tilde R_\xi(q) (1 - e^{- 2 \nu q^2 t} )\,.
\eea 
Splitting this term in two pieces and integrating each of them separately over time, we obtain
\bea
\overline{h} = 
\frac{\lambda_0 R(0)}{4 \nu} t  - \frac{\lambda_0}{4 \nu}  \int \frac{dq}{2\pi} \tilde R_\xi(q) \frac{1 - e^{- 2 \nu q^2 t} }{2 \nu q^2}\,.
\eea 
One recognizes the same integral entering the second moment, and so  for $t \ll t^*$ we have
\bea
\overline{h} = v_0 t - \frac{\lambda_0}{4 \nu} \overline{h^2} \,,
\eea 
which indeed reproduces exactly, for $t \gg t_f$, the leading {\it negative} correction in Eq. (\ref{log1}). 
The first term linear in time is however always present, and non-universal. 
The same exact term arises in the DP, and corresponds to the multiplicative contribution to the
moments
$\overline{Z^n} = e^{t n \frac{1}{2 T^2} R_V(0)} \equiv e^{t n \frac{\lambda_0^2}{8 \nu^2} R_\xi(0)}$ 
arising from the equal replica (self-energy) terms after averaging the replicated partition sum. 
Although usually dropped, these terms are present and correspond to an additive 
(non-universal) non-random correction $\frac{\lambda_0^2}{8 \nu^2} R_\xi(0) t$ to $\ln Z$.


\end{document}